\documentclass[twocolumn,preprintnumbers,amsmath,amssymb,prl,superscriptaddress,showpacs]{revtex4}

\usepackage{graphicx}
\usepackage{dcolumn}
\usepackage{bm}

\begin{document}

\title{Theoretical prediction of rotating magnon wavepacket in ferromagnets} 

\author{Ryo Matsumoto}

 \affiliation{Department of Physics, Tokyo Institute of Technology, Tokyo 152-8551, Japan}

\author{Shuichi Murakami}%
\email{murakami@stat.phys.titech.ac.jp}
 \affiliation{Department of Physics, Tokyo Institute of Technology, Tokyo 152-8551, Japan} 
\affiliation{
PRESTO, Japan Science and Technology Agency (JST), Saitama, Kawaguchi 
332-0012, Japan }


\date{\today}

\begin{abstract}
We theoretically show that the magnon wavepacket has a rotational motion in two ways; 
a self-rotation and a motion along the boundary of the sample (edge current). 
They are similar to cyclotron motion of electrons, but unlike electrons the magnons 
have no charge and the rotation is not due to Lorenz force.
These rotational motions are caused 
by the Berry phase in momentum space from magnon band structure. 
Furthermore, these rotational motions of the magnon give an additional correction term to the magnon Hall effect. 
We also discuss the Berry curvature effect in the classical limit of long-wavelength magnetostatic spin waves having 
macroscopic coherence length. 
\end{abstract}

\maketitle

\textit{Introduction}---
Spin wave (magnon) in an insulating magnet is a low-energy collective excitation~\cite{Kittel, Hillebrands}.  
It has recently been focused as a tool for the spintronics application, because
it can have a good coherence, compared with the spin current in metals. 
The motions of the magnons are now measurable in a time- and space-resolved way with reasonable accuracy.
It can be experimentally generated and detected via the spin Hall effect \cite{Kajiwara},
and for the spintronics application, a precise spatial and temporal control of spin wave is 
desired. 

\begin{figure}
\includegraphics[scale=0.6]{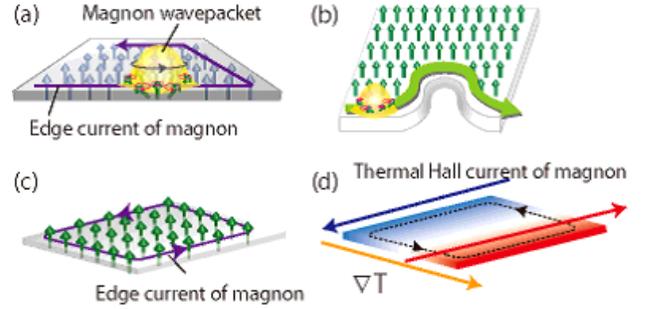}
\caption{(Color online) (a) Self-rotation of a magnon wavepacket and a magnon edge current. 
(b)  
The magnon near the boundary proceeds along the boundary, irrespective of
the edge shape. 
(c) Magnon edge current in equilibrium. (d) Under the temperature gradient, the amount of the transverse heat current are not balanced between the two edges, and a finite thermal Hall current will appear. }
\label{fig magnon Hall effect}
\end{figure}

In the present paper, we theoretically find that the motion of magnon wavepackets 
in insulating magnets undergoes a self-rotational motion and a 
rotational motion along the edge of the sample (Fig.~\ref{fig magnon Hall effect}(a)). 
The latter motion gives rise to the thermal Hall effect of magnons~\cite{Katsura, Onose, Fujimoto}. 
This 
phenomenon is due to the Berry curvature in momentum space, representing the topological structure 
in the magnon bands. 
We theoretically predict that if a magnon wavepacket is excited in the vicinity of the edge of a sample, it will move along the edge. 
It is expected to be visible in some magnets having macroscopic coherence length 
($\sim 10$mm)~\cite{Buttner}, and it should be a powerful tool for exploring effects of Berry phase in momentum space.

We calculate the transverse thermal transport coefficients for a magnon system by two methods: the semiclassical theory and the linear response theory, by analogy with an electron system. We show that both theories give the same result and the thermal Hall conductivity $\kappa^{xy}$ can be written in terms of the Berry curvature. 
Our theory includes the contribution of magnon rotational motion, which has been overlooked in the previous
 theory of magnon thermal Hall effect \cite{Katsura,Onose}.
From this we show that the thermal Hall effect of the magnon arises from 
the edge current of the magnon.   
We apply our theory to various magnons, including both the exchange spin wave (quantum-mechanical magnon) 
e.g. in a ferromagnet Lu$_2$V$_2$O$_7$, and the magnetostatic spin wave in yttrium-iron-garnet (YIG) films.
The results for thermal Hall conductivity in Lu$_2$V$_2$O$_7$ roughly reproduces the experiment \cite{Onose}. 
Throughout this paper we consider localized spin systems on a two-dimensional lattice, and assume that there is no interaction between magnons. 

\textit{ Semiclassical theory} ---
The dynamics of a wavepacket of electrons in a periodic system can be described by the semiclassical equation, including the topological Berry phase effect \cite{Sundaram Niu}. 
When a force is exerted 
onto the electron, there occur various intrinsic Hall effects due to the Berry phase.
In analogy with this, we construct the semiclassical equation of motion of magnons. We consider a wavepacket of a magnon which is localized both in real- and momentum-space. If there exists a slowly-varying potential $U(\bm{r})$ for the magnons, they feel a force. Following \cite{Sundaram Niu,Xiao Niu}, we derive the semiclassical equations of motion for the magnon wavepacket as:
\begin{align}
	\dot{\bm{r}}_n	&=	\frac{1}{\hbar}\frac{\partial \varepsilon_{n\bm{k}} }{\partial \bm{k}}-\dot{\bm{k}}\times\bm{\Omega}_n(\bm{k}), 
\ \ 
\hbar \dot{\bm{k}}	=	-\nabla U(\bm{r}) \label{eq eomk},
\end{align}
where $\bm{\Omega}_n(\bm{k})$ is the Berry curvature: $\bm{\Omega}_n(\bm{k})=i\left\langle \frac{\partial u_n}{\partial \bm{k}}\right|\times\left| \frac{\partial u_n}{\partial \bm{k}}\right\rangle$ with $ | u_n(\bm{k}) \rangle$ being the periodic part of the Bloch waves in the $n$th band and $\varepsilon_{n\bm{k}}$ is the $n$th band energy of magnon.  
For the potential $U(\bm{r})$ we cannot use the electric field 
because the magnons have no charge. 
Instead we focus on the boundary of the system, which can be regarded as a confining 
potential $U({\bf r})$.  
On the edge along $y$-direction, for example, the gradient of the confining potential $\partial_x U(\bm{r})$ produces 
an anomalous velocity 
$\dot{\bm{k}}\times\bm{\Omega}_n=-\hbar^{-1}\partial_x U(\bm{r})\Omega_{n,z}(\bm{k})\hat{y}$ in Eq.~\eqref{eq eomk}. 
By summing it over all the occupied states, we get an edge current  $I_y=	-\int_a^b dx
 \partial_xU(\bm{r})\left( \frac{1}{\hbar V}
\sum_{n,\bm{k}} \rho(\varepsilon_{n\bm{k}} +U(\bm{r}))\Omega_{n,z}(\bm{k})	 \right)=	-\frac{1}{\hbar V} \sum_{n,\bm{k}} \int_{\varepsilon_{n\bm{k}}}^{\infty} d\varepsilon 
\rho(\varepsilon) \Omega_{n,z}(\bm{k})$,  
where $x=a$ and $x=b$ represent the inside and the outside of the system, respectively, such that $U(a)=0$ and $U(b)=\infty$, $\rho(\varepsilon )$ is the Bose distribution function $\rho(\varepsilon )=(e^{(\varepsilon  -\mu)/k_{\text{B}}T}-1)^{-1}$, $k_{\text{B}}$ is the Boltzmann constant, $\mu$ is the chemical potential and $T$ is the temperature. 
Since the magnon edge current is independent of the edge direction, we simply write this as $I$. 
This magnon edge current $I$ is independent of 
the form of the confining potential. 
Therefore, the edge current circulates along the whole edge (Fig.~\ref{fig magnon Hall effect}(b)).
Strictly speaking, this approach is applied only when $U$ is slowly varying. Nevertheless, as $I$ 
does not depend on the form of $U(\bm{r})$, this remains valid even when $U(\bm{r})$ is representing 
a hard wall and rapidly varying. Such kind of approach has been successful in quantum Hall systems \cite{Buttiker}.
We can therefore expect that the similar approach is successful also for magnons; we can use 
the slowly varying confining potential which is much easier for theory in order to predict 
physical phenomena occuring irrespective of the details of the potential. 
We note that if the coherence length is short, the effective system size is given by the coherence
length of the magnon, and the edge current is also confined within this length scale. 

If either 
$\mu$ or $T$ varies spatially, this 
otherwise circulating current will no longer cancel in the 
interior of the system. 
This causes the thermal Hall effect as we show in the following. 
The magnon current and energy current due to the edge current can then be written respectively as 
$\bm{j}=\nabla \times \frac{1}{\hbar V} 
\sum_{n,\bm{k}} 
\int_{\varepsilon_{n\bm{k}}}^{\infty} \rho(\varepsilon )\bm{\Omega}_n(\bm{k}) d\varepsilon $, 
$\bm{j}_{\text{E}}=\nabla \times \frac{1}{\hbar V} 
\sum_{n,\bm{k}} 
\int_{\varepsilon_{n\bm{k}}}^{\infty}\varepsilon \rho(\varepsilon )\bm{\Omega}_n(\bm{k})    d\varepsilon
$. 
Various thermal coefficients are derived from these equations. 
If the chemical potential $\mu$ or the temperature $T$ spatially varies, the magnon current and heat current is induced 
via the Bose distribution function.
For instance, in the presence of a temperature gradient in the $y$ direction, the magnon current in the $x$ direction are written as $\left( j\right)_x^{\nabla T} 	= T \partial_y \left( \frac{1}{T} \right)   \frac{1}{\hbar V}
\sum_{n,\bm{k}} 
\int_{\varepsilon_{n \bm{k} } }^{\infty}  (\varepsilon - \mu)  \left(\frac{d\rho}{d\varepsilon} \right) \Omega_{n,z} (\bm{k})d\varepsilon$.
Other thermal transport coefficients can be obtained in the same way. 
Now we write the linear response of the magnon current and the heat current as
\begin{eqnarray}
	\bm{j}	&=L_{11}\left[ -\nabla U	
-\nabla \mu 
\right]
              + L_{12}\left[ T \nabla\left(\frac{1}{T}\right) \right]  ,\\	
	\bm{j}_{\text{Q}}	&=L_{12}\left[ -\nabla U
-\nabla \mu  \right] 
                 + L_{22}\left[ T \nabla\left(\frac{1}{T}\right)  \right], \end{eqnarray}
where the heat current $\bm{j}_{\text{Q}}$ is defined as  $\bm{j}_{\text{Q}} \equiv \bm{j}_{\text{E}}-\mu\bm{j}$. 
We take $\mu=0$ here, because the magnon number is not conserved.
The transport coefficients can be written as 
\begin{equation}
L^{xy}_{ij}=-\frac{(k_{\text{B}}T)^q}{\hbar V}
\sum_{n,\bm{k}}
\Omega_{n,z}(\bm{k})c_{q}(\rho_n),
\end{equation}
where $V$ is the area of the system, $\rho_n\equiv \rho(\varepsilon_n(\bm{k}))$,  
$c_q(\rho)
=\int_0^{\rho} \left( \log (1+t^{-1}) \right)^q dt
$, $q=i+j-2$ with $i,j=1,2$. 
For example, $c_0(\rho)=\rho$, 
$c_1(\rho)=   \left( 1+\rho   \right) \log  \left( 1+\rho  \right) -\rho  \log \rho  $, and 
$c_2(\rho) =   \left( 1+\rho   \right) \left( \log   \frac{1+\rho}{\rho}  \right)^2 - \left(  \log \rho \right)^2 -2\text{Li}_2(-\rho) $, where $\text{Li}_2(z)$ is the polylogarithm function. 
Thus the thermal Hall conductivity $\kappa^{xy}={L_{22}^{xy}}/T$
can be obtained as 
\begin{equation}
\kappa^{xy}=-\frac{k^2_{\text{B}}T}{\hbar V}\sum_{n,\bm{k}} c_2(\rho_n) 
\Omega_{n,z}(\bm{k}). 
\label{eq kappaxy2}
\end{equation}
From Eq.~\eqref{eq kappaxy2} we can see that $\kappa^{xy}$ comes from the 
Berry curvature ${\bf \Omega}_{n}({\bm{ k}})$ in momentum space. Therefore, if the energy band is close to each other, i.e. near the 
band crossing, this gives a large contribution to $\kappa^{xy}$. 

Thus, the thermal Hall conductivity solely comes 
from the edge magnon current.
In equilibrium (Fig.~\ref{fig magnon Hall effect}(c)), there exists the edge current of the magnon due to the confining potential gradient. This magnon edge current circulates along the boundary, giving no net thermal current across 
the magnet. If the temperature gradient is applied (Fig.~\ref{fig magnon Hall effect}(d)), the balance of the contributions to the heat current from the two opposite edges will be broken, and finite thermal Hall current will appear. 
Because the 
temperature gradient is a statistical force, it neither exerts a force
to the magnons nor deflects the wavepacket in the bulk. 
The  edge current should be observed experimentally by using time and space resolved Brillouin light scattering technique~\cite{Hillebrands}.


\textit{Linear response theory}---
There is a discrepancy between Eq.~(\ref{eq kappaxy2}) and the 
formula obtained in \cite{Katsura,Onose} using the Kubo formula.
In \cite{Katsura,Onose}, the term from magnon rotational motion is 
missing, as we discuss in the following. 

Because the temperature gradient is not a dynamical force directly 
acting onto the particle, but a statistical force, 
its theoretical treatment requires some care.
In the linear response theory for electronic systems including heat current 
and temperature gradient, it is convenient to introduce a fictitious 
gravitational field $\psi$~\cite{Luttinger}, which exerts a force to 
the wavepacket, proportional to its energy.
Such formalism can be applied to the magnon system, taking into 
consideration several differences that the magnon 
has no charge and is not a fermion but a boson. 
As a result,
the transport coefficients for magnons
consist of  two terms:  
one term from a deviation of a particle density operator from an equilibrium calculated by the Kubo formula, 
and the other term 
arising from the deviations of current operators which are linear in applied fields. 
The latter term is expressed in term of the reduced orbital 
angular momentum of magnons $\sim \langle \bm{r}\times\bm{v}\rangle$, but was missing in \cite{Katsura,Onose}.
This latter term has been discussed in the context of orbital magnetization in 
electron systems \cite{Smrcka Streda, Oji Streda, Bergman, Xiao1, Xiao2, Thonhauser}. 
The calculation is in parallel with that for electrons \cite{Smrcka Streda, Oji Streda}, 
and its details will be presented elsewhere. 
As a result, 
the thermal Hall conductivity from the linear response theory is identical with that from the semiclassical theory in
Eq.~\eqref{eq kappaxy2}. 
The new term to the linear response theory corresponds to the reduced orbital angular momentum of magnon.  
It consists of two parts: the edge current and the self-rotation of the wavepacket.
The reduced angular momentum for the edge
current per unit volume is 
\begin{align}
l^{\text{edge}}_z=-\frac{2}{\hbar V}
\sum_{n,\bm{k}} 
\int_{\varepsilon_{n\bm{k}}}^{\infty}d\varepsilon 
\rho(\varepsilon)\Omega_{n,z}(\bm{k}), \label{eq L edge}
\end{align}
and that for the self-rotation is calculated in analogy with the electron system \cite{M C Chang} as, 
\begin{eqnarray}
l_z^{\text{self}}	=	-\frac{2}{\hbar V} \text{Im} \sum_{n,\bm{k}}  \rho_n \left\langle \left. \frac{\partial u_n}{\partial k_x} \right|  \left( H-\varepsilon_{n\bm{k}} \right)   \left|  \frac{\partial u_n}{\partial k_y} \right. \right\rangle  \label{eq L self}.
\end{eqnarray} 
Namely,  
in addition to this edge current, we find that the magnon wavepacket rotates around  
itself and induces orbital angular momentum, because of the Berry phase in momentum space. 
Thus the magnon in equilibrium has in
general a nonvanishing orbital angular momentum due to the Berry curvature. 
This magnon orbital
motion can be regarded as a generalized cyclotron motion, 
whereas the magnon 
feels no Lorentz force and cannot have a cyclotron motion
in the same sense as that of electrons. In this respect, this motion 
is purely due to the magnon band structure. This effect is common in various 
wave phenomena like electrons\cite{Xiao Niu}, 
photons \cite{Onoda}, and so forth.

Orbital motions of electrons give rise 
to a magnetic moment due to the electron charge. On 
the other hand, magnons have no charge, but have a magnetic moment. 
Because the magnon carries magnetic dipole, the rotating magnon 
wavepacket can be regarded as a circulating spin current. 
Hence as is similar to the spin Hall effect, and its 
insulator counterpart, i.e.~the magnetoelectric effect in noncollinear 
spin structure \cite{KatsuraME}, the rotating magnon wavepacket 
should accompany a polarization charge. 
It requires the spin-orbit interaction, 
 i.e., Dzyaloshinskii-Moriya (DM) interaction.
This effect is dual to the rotation of electric charge, producing 
a magnetic dipole. 

\begin{figure}
\includegraphics[scale=0.5]{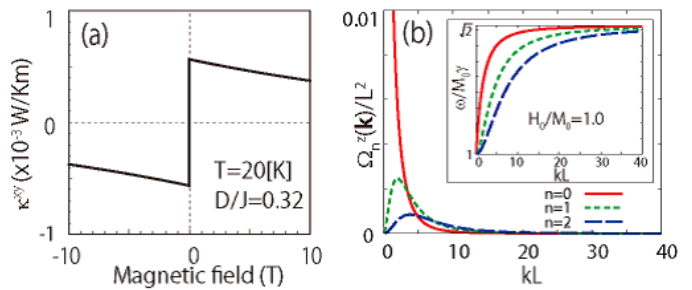}
\caption{(Color online) Numerical results. (a)  
thermal Hall conductivity of the magnon in Lu$_2$V$_2$O$_7$ in the magnetic field at $T=20[\text{K}]$. 
(b) Berry curvature $\Omega_n^z(k)$ with $n=0,1,2$ for the magnetostatic spin waves in YIG. 
Eigenmode dispersions are shown in the inset. }
\label{fig kappaxy}
\end{figure}

\textit{Thermal Hall effect of $\text{Lu}_2\text{V}_2\text{O}_7$ }---
Now we apply our results to the ferromagnetic Mott-insulator $\text{Lu}_2\text{V}_2\text{O}_7$ with a pyrochlore structure. This material has spin-1/2 $\text{V}^{4+}$ ions with the DM interaction. The collinear ferromagnetic ground state is stable because the total DM vectors of the bond sharing the same site is zero~\cite{Onose}, and the effective spin-wave Hamiltonian is written as 
$H_{\text{eff}}=\sum_{\langle i,j \rangle}-J\bm{S}_i \cdot \bm{S}_j + \bm{D}_{ij} \cdot \left( \bm{S}_i \times \bm{S}_j \right) -g\mu_{\text{B}}\bm{H}\cdot \sum_i\bm{S}_i$,
where $\langle i,j \rangle$ denotes the nearest neighbor pairs, $J$ is the exchange interaction, $\bm{D}$ is the DM vector, $g$ is the g-factor, $\mu_{\text{B}}$ is Bohr magneton, and $\bm{H}$ is the magnetic field in the $z$ direction. 
We focus on the temperature regime much lower than the Curie temperature $T_C=70[\text{K}]$, 
for existence of well-defined Bloch waves of magnons. We can then assume that the contribution from the lowest band is dominant, whose Berry curvature is 
$\Omega_{1,z}\simeq -\frac{A^4}{8\sqrt{2}}\frac{D}{J}\frac{H_z}{H}(k_x^2+k_y^2+2k_z^2)$ as calculated in \cite{Onose},
with $A$ being a quarter of the lattice constant. 
Using this, we can estimate the orbital angular momentum of the magnon from both the self-rotation motion $L_z^{\text{self}}$  and the edge current $L_z^{\text{edge}}$. 
Near $k=0$, the lowest-band dispersion is quadratic 
and we can introduce the effective mass of the magnon of the lowest band $m_1^*$, defined as $m^*_n\equiv \hbar^2(\partial^2 \varepsilon_{n\bm{k}}/\partial k^2)^{-1}$. 
The orbital angular momentum from the self-rotation motion of the magnon per unit volume is calculated analytically, 
\begin{equation}
L_z^{\text{self}}=-\frac{JSm_1^*}{\hbar A}  \frac{D}{J} \frac{1}{32\pi^{3/2}}\left(\frac{k_\text{B}T}{JS}\right)^{5/2} \text{Li}_{\frac{5}{2}}\left(e^{-\frac{k_\text{B}T}{ g \mu_{\text{B}} H}}\right). \label{eq Lz self LuVO} 
\end{equation}
We obtain $L_z^{\text{self}} \simeq -0.009 \hbar$ and $L_z^{\text{edge}} \simeq +0.008\hbar $ per unit cell.
We can also calculate the thermal Hall conductivity $\kappa^{xy}$, 
assuming that the contribution of the lowest band is dominant. 
The resulting $\kappa^{xy}$ is shown in Fig.~\ref{fig kappaxy}(a), which roughly agrees with the experimental results 
\cite{Onose}.

\textit{Magnetostatic spin waves} ---
Our theory is also applied to the magnetostatic spin waves in a ferromagnet. 
In the magnetostatic spin waves, 
the wavelength are sufficiently long and magnetic anisotropy is mostly determined by the demagnetizing field
which is dependent on the sample shape. The exchange coupling is negligible because of the long wavelength. 
This anisotropy due to the demagnetizing field plays the similar role as the spin-orbit coupling, and gives
rise to the Berry curvature contributing to the Hall effect of spin waves. As an example, we 
consider YIG films, magnetized to the saturation in an arbitrary direction by an external magnetic field.
YIG is a ferrimagnetic insulator, and the spin wave in YIG can propagate over centimetres. 
The spin wave mode is expressed as a plane wave: 
$\bm{m}(\bm{r},t)=\bm{m}(z)\exp(i(\bm{k}\cdot\bm{r}_{\|}-\omega t))$, where $\bm{m}$ is a two-dimensional vector 
perpendicular to the saturation magnetization $\bm{M}_{0}$, $z$ is a coordinate perpendicular to the film, 
$\bm{r}_{\|}$ is the coordinate within the film,  
and $\omega$ is a frequency of the spin wave. 
The linearized Landau-Lifshitz equation, coupled with the Maxwell equation with boundary conditions for the film, 
is cast into the integral equation~\cite{Kalinikos}:
$ \omega_{H} \bm{m}(z) -\omega_M \int_{-L/2}^{L/2}dz^{\prime} \hat{G}(z,z^{\prime}) \bm{m}(z^{\prime})=\omega \sigma_y \bm{m}(z). 
$
Here, $\omega_H=\gamma H_0$, $\omega_M=\gamma M_0$, $L$ is the film thickness,  $\hat{G}(z,z^{\prime})$ is $2\times2$ complex matrix of the Green's function defined
in \cite{Kalinikos}, $\sigma_y$ is the Pauli matrix, 
$\gamma$ is the gyromagnetic ratio, $H_0$ is a static magnetic field. 

This integral equation is a generalized eigenvalue problem due to the presence of $\sigma_y$. The 
calculation of the Berry curvature then requires some modifications.
Following the prescription of the 
wavepacket dynamics~\cite{Chuanwei}, we introduce the Berry curvature of the magnetostatic spin wave 
$\Omega_n^{\gamma}(\bm{k})=-\epsilon_{\alpha\beta\gamma}\text{Im}\left\langle \frac{\partial \bm{m}_{n,\bm{k}}}{\partial k_{\alpha}}  \right|   \sigma_y  \left|       \frac{\partial \bm{m}_{n,\bm{k}}}{\partial k_{\beta}} \right\rangle ,
$
where $\epsilon_{\alpha\beta\gamma}$ is the totally antisymmetric tensor, $n$ is a band index of the spin wave, 
and the bra-ket product refers to a usual inner product of vectors and an integral over $z$. 
%
In some cases, this Berry curvature vanishes because of symmetry. 
This occurs when $\bm{M}_{0}$ lies in the plane. 
In this case the system is invariant under
product of the time-reversal operation and the $\pi$ rotation within the film, and 
this symmetry forces the Berry curvature $\Omega_n(\bm{k})$ to be zero. 
%
On the other hand, when $\bm{M}_{0}$ is not in the plane, the Berry curvature is expected to be nonzero for any modes, 
leading to the 
rotational motion of the wavepackets and the Hall effect. Actually, if $\bm{M}_{0}$ is perpendicular to the plane, i.e., for the magnetostatic forward volume wave (MSFVW) mode, we can calculate the Berry curvature. From Ref.~\cite{Damon}, 
the frequency $\omega=\omega_n$ for $n$th eigenmode  is determined by $\sqrt{p}\tan(\frac{\sqrt{p}kL}{2}+\frac{n\pi}{2})=1$ 
where
$p=\frac{\omega_M\omega_H}{\omega^2-\omega_H^2}-1$ and $n=0,1,2,\cdots$, and is shown in 
the inset of Fig.~\ref{fig kappaxy}(b) for $n=0,1,2$.
The modes are 
\begin{equation}
\bm{m}_{n\bm{k}}(z)=
\frac{\omega_M\cos\left(
\sqrt{p}kz+\frac{n\pi}{2}\right)}{\sqrt{N}(\omega_H^2-\omega_n^2)}(i\omega_H\bm{k}-
\omega(\hat{z}\times\bm{k})),
\end{equation} where  $N$ is the normalization constant
determined by $\left\langle \bm{m}_{n,\bm{k}}  \right|   \sigma_y  \left|   \bm{m}_{n,\bm{k}} 
\right\rangle=1$. We then obtain for $n$-th mode;
\begin{equation}
\Omega_n^{z}(\bm{k})=\frac{1}{2\omega_H}\frac{1}{k}\frac{\partial \omega_n}{\partial k} \left( 1-
\frac{\omega_H^2}{\omega_n^2}\right). 
\end{equation}
$\Omega_n^z$ is evaluated numerically, and the results are shown in Fig.~\ref{fig kappaxy}(b).
We have thus confirmed that the Berry curvature is nonzero for the MSFVW mode, and the 
rotations of magnon wavepacket are predicted to occur. 

\textit{Conclusions}---
To summarize, we found that the magnon wavepacket undergoes rotational motions in two ways: self-rotation and motion along the edge. The latter is responsible for the magnon thermal Hall effect. These rotational motions are due to the Berry phase, and is similar to the electron cyclotron motion, but without a Lorentz force. 
The present theory  
is applied both to the exchange spin wave (quantum-mechanical magnon) e.g.~in 
Lu$_2$V$_2$O$_7$, and to the classical 
magnetostatic waves e.g.~in YIG. This effect is expected to be observed via space- and time-resolved observation of the magnon wavepacket with macroscopic coherence length.

Similarly to 
 anomalous Hall or spin Hall effect, the Berry curvature in the
magnon systems is enhanced near band crossings, where the magnon frequency in a focused band is close to that of other 
bands. One can design the magnonic crystals~\cite{magnonics} so that the magnon bands have a band crossing.
One can then expect to have a prominent rotational motions of magnon wavepackets, if its wavenumber is close to the 
band crossing.

We would like to thank B.~I.~Halperin and T.~Ono for discussions.
This work is partly supported  
by Grant-in-Aids  
from MEXT, Japan 
 (No.~21000004 and 22540327), 
and by the Global
Center of Excellence Program by MEXT, Japan through the
"Nanoscience and Quantum Physics" Project of the Tokyo
Institute of Technology.

\end{document}